\documentclass[onecolumn,preprintnumbers,amsmath,rmp]{revtex4}

\usepackage{graphicx}
\usepackage{epsfig}

\begin{document}
\title{\bf Radiative Correction to The Casimir Energy with Mixed Boundary Condition in $2+1$ Dimensions}
\author{M A Valuyan}
\email{m-valuyan@sbu.ac.ir; m.valuyan@semnaniau.ac.ir}
\affiliation{Department of Physics, Semnan Branch, Islamic Azad University, Semnan, Iran}
\affiliation{Energy and Sustainable Development Research Center, Semnan Branch, Islamic Azad University, Semnan, Iran}
\date{\today}
\vspace{1cm}
\begin{abstract}\textbf{ABSTRACT}\par
In the present study, the zero- and first-order radiative correction to the Casimir energy for the massive and massless scalar field confined with mixed\,(Neumann-Dirichlet) boundary condition between two parallel lines in $2+1$ dimensions for the self-interacting $\phi^4$ theory was computed. The main point in this study is the use of a special program to renormalize the bare parameters of the Lagrangian. The counterterm used in the renormalization program, which was obtained systematically position-dependent, is consistent with the boundary condition imposed on the quantum field. To regularize and remove infinities in the calculation process of the Casimir energy, the Box Subtraction Scheme as a regularization technique was used. In this scheme, two similar configurations are usually introduced, and the vacuum energies of these two configurations in proper limits are subtracted from each other. The final answer for the problem is finite and consistent with the expected physical basis. We also compared the new result of this paper to the previously reported results in the zero- and first-order radiative correction to the Casimir energy of scalar field in two spatial dimensions with Periodic, Dirichlet, and Neumann boundary conditions. Finally, all aspects of this comparison were discussed.\vspace{0.5cm}\\
\emph{Keywords:} Casimir energy, Renormalization, Boundary condition, Regularization.\\
PACS Nos: \emph{11.10.−z; 11.10.Gh; 11.25.Db; 11.15.Bt}
\end{abstract}
\maketitle

\section*{1. Introduction}
\label{sec:intro}
Since conducting the first study in a radiative correction to the Casimir energy by Bordag et al., various studies in this category of the problem have been conducted\,\cite{Bordag.et.al.1,Bordag.et.al.2,Bordag.et.al.3}. The diversity of these studies was expressed in several sections. In the literature, radiative correction to the Casimir energy has been conducted for various quantum fields under different boundary conditions\,\cite{quantum.field.theory.1}. In addition, this quantity in some curved manifolds was also obtained previously\,\cite{curved.valuyan.1,curved.valuyan.2}. One of the main issues in these sorts of the problem is to determine what types of counterterms would be suitable in the renormalization program\,\cite{reza.Eurpjc}. In the earlier works, considering the type of boundary conditions imposed on quantum fields does not affect selection of the type of the counterterms in the renormalization program. Therefore, the \emph{free counterterm} has been used in any problem with any boundary condition imposed\,\cite{RC.free.counterterms.1,RC.free.counterterms.2}. Later, in some works, the influence of nontrivial boundary conditions was imported in renormalization programs\,\cite{5050RCfree.notfree.1}. However, this importing process was not performed completely. Indeed, the authors applied free counterterms in the space between the boundaries and placed additional surface counterterms at the boundaries\,\cite{5050RCfree.notfree.2}. Gousheh et al. implemented an integrated scheme to consider all influences of the boundary condition imposed on the quantum fields in the renormalization program\,\cite{SS.Gousheh.etal.1,SS.Gousheh.etal.2}. In their scheme, a systematic perturbation expansion was performed, and after applying the renormalization condition, a \emph{position-dependent counterterm} was obtained. Similar to the free counterterm, the position-dependent counterterms were obtained from the $n$-point function in the usual perturbation theory, except that for this type of counterterms, the relevant Green's function was used in every needed places. This employment of the Green's function caused the counterterms to be position-dependent, and the effect of the boundary conditions was involved in the counterterms. We maintain that, when a boundary condition is applied to the quantum fields, the renormalization program and all of its contributed elements like counterterms should be affected owing to the dominant boundary conditions. Therefore, the use of a unique counterterm\,(free counterterm), regardless of the type of the boundary condition imposed on the quantum fields for any problem, may not be legitimate. Moreover, the counterterms are forced to renormalize the divergent contribution of the bare parameters of the system. Therefore, if it has not been chosen properly, it may cause some divergent physical quantities. The radiative correction to the Casimir energy for $\phi^4$ theory confined between two parallel lines with Dirichlet boundary conditions in $2+1$ dimensions was obtained as a divergent result\,\cite{cavalcanti.1,cavalcanti.2}. This quantity, by changing the counterterms to position-dependent ones, which is consistent with the dominant boundary conditions, is to be convergent\,\cite{2dim.valuyan}. The use of the position-dependent counterterms in the calculation of the radiative correction to the Casimir energy for different geometries with various quantum fields and boundary conditions was examined in\,\cite{my.other.paper.1,my.other.paper.2,my.other.paper.3}. Application of this type of counterterm in the radiative correction to the Casimir energy on some curved manifolds is also successful\,\cite{curved.valuyan.1,curved.valuyan.3}. The obtained results for all cases are consistent with the expected physical basis. In this study, using the position-dependent counterterms, we calculated the radiative correction to the Casimir energy for the massive and massless scalar fields confined between two parallel lines with mixed\,(Dirichlet-Neumann) boundary conditions. This calculation was earlier performed using the free counterterms in the renormalization program\,\cite{cavalcanti.1,cavalcanti.2}. However, we solved this problem again by employing the position-dependent counterterms, and found that our result differed from those reported previously\,\cite{cavalcanti.2}.
\begin{figure}[th] \hspace{0cm}\includegraphics[width=7cm]{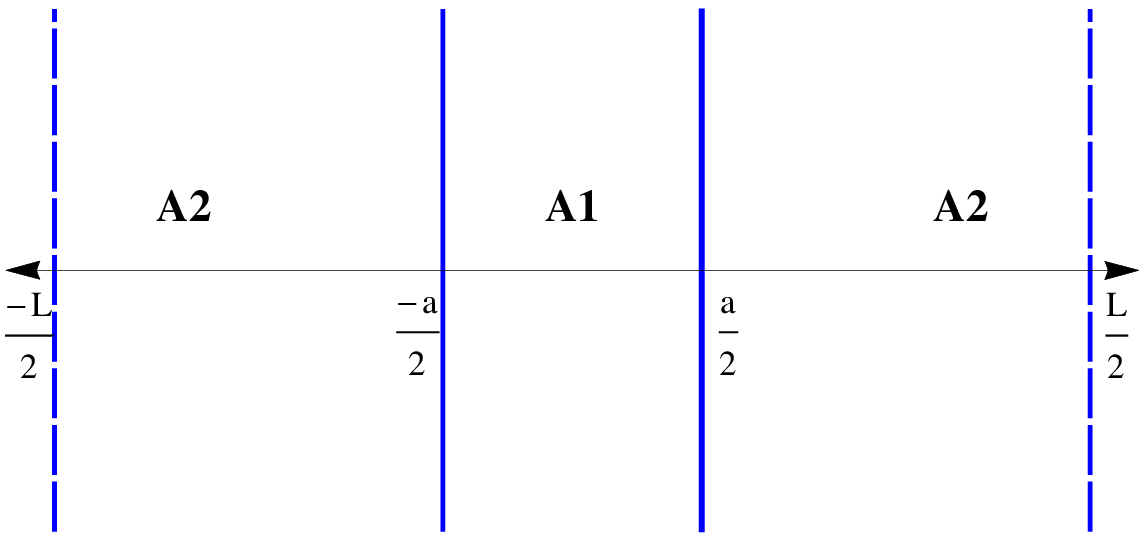}\hspace{1.3cm}\includegraphics[width=7cm]{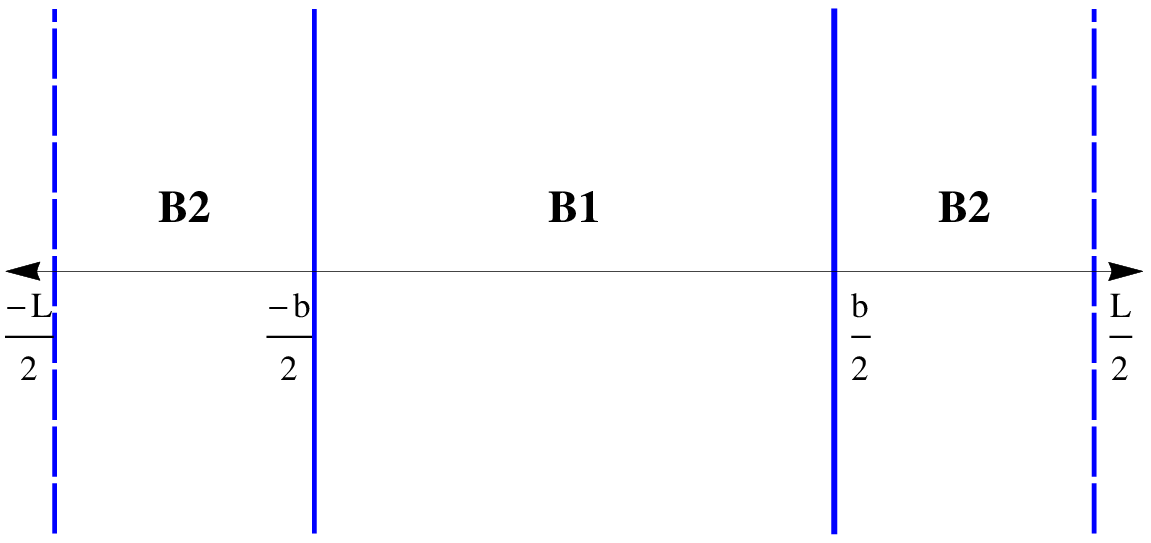}\caption{\label{BSS.Fig}
  The left figure is ``$A$ configuration" and the right one is ``$B$
  configuration".}
\end{figure}
\par
The main part of the computation of the Casimir energy is to deal with divergent expressions. To illustrate that what kinds of the divergences appear in the calculation and how can be removed, several regularization techniques were introduced in the literature\,\cite{quantum.field.theory.1,quantum.field.theory.2,quantum.field.theory.3,RC.free.counterterms.2} and the merit or ability of these regularization techniques was discussed. Some of the known important regularization techniques in this subject are Mode summation technique, Zeta function regularization technique, Green's function formalism, and the Box subtraction scheme\,\cite{mode.summation.1,mode.summation.2,mode.summation.3,Green.function.1,Green.function.2,Green.function.3}. In this study, to regularize the infinities appeared in the calculation of the radiative correction to the Casimir energy, we used the Box Subtraction Scheme\,(BSS). Through the BSS, two similar configurations were defined, and the vacuum energies of these two configurations were subtracted from each other\,\cite{boyer.,BSS.man.1,BSS.man.2}. Fig.\,(\ref{BSS.Fig}) displays these two configurations for our problem. The Casimir energy can now be defined as:
\begin{eqnarray}\label{BSS.Def.}
    E_{\mbox{\tiny Cas.}}=\lim_{L/b\to\infty}\lim_{b/a\to\infty}\big[E_{\mathcal{A}}-E_{\mathcal{B}}\big],
\end{eqnarray}
where $E_{\mathcal{A}}-E_{\mathcal{B}}=E_{A1}+2E_{A2}-E_{B1}-2E_{B2}$. As fig.\,(\ref{BSS.Fig}) shows, $E_{\mathcal{A}}$ and $E_{\mathcal{B}}$ are the total vacuum energies of configurations $A$ and $B$, respectively. $E_{A1}$, $E_{A2}$, $E_{B1}$ and $E_{B2}$ display the vacuum energy of each region separated by lines in fig.\,(\ref{BSS.Fig}). Considering the Casimir energy problem designed in the even spatial dimensions, are usually accompanied by some degrees of complexity. This complexity commonly originates from the kind of divergences appearing in the vacuum energy, since the type of divergent expression appeared in the vacuum energy of systems in even spatial dimensions is usually logarithmic, and the removal process of this type of divergence is more difficult than that of the other types of divergence. The successful experience of the aforementioned renormalization program supplemented by BSS as a regularization technique in \,\cite{2dim.valuyan} motivated us to calculate the radiative correction to the Casimir energy for a massive and massless scalar field with mixed boundary conditions in $2+1$ dimensions. The remainder of this work is organized as follows. In Section 2, radiative correction to the Casimir energy between two parallel lines in two spatial dimensions for massive and massless scalar fields is computed. The obtained results are consistent with the expected physical ground. In the next section, based on the previously reported results in the Casimir energy under Dirichlet, Neumann, and Periodic boundary conditions in $2+1$ dimensions, the aspects of results are discussed.

\section*{2. First-Order Radiative Correction to the Casimir Energy}
\label{sec:RC.Cas.Cal.}
In this section, we compute the first-order radiative correction to the Casimir energy for the self-interacting massive scalar field\,($\lambda\phi^4$ theory) confined with mixed boundary condition between two lines in two spatial dimensions. For this purpose, the vacuum energy expression up to the first-order of coupling constant $\lambda$ is required. Since the calculation and deduction of the vacuum energy using the position-dependent counterterms were reported previously in the literature, we do not mention the details of this computation here\,(for more details see Refs.\,\cite{reza.Eurpjc,my.other.paper.3}). Using their final results, the total vacuum energy expression up to the first-order of coupling constant $\lambda$ becomes:
\begin{eqnarray}\label{1st.order.vac.energy}
  E^{(1)}=\frac{-\lambda}{8}\int G^2(r,r)d^2\mathbf{x},
\end{eqnarray}
where the coordinates $r=(t,\mathbf{x})$ and $\mathbf{x}=(x,y)$. Additionally, $G(r,r)$ is the Green's function and the superscript $(1)$ denotes the first-order in the coupling constant $\lambda$. To calculate the Casimir energy according to eqs.\,\eqref{BSS.Def.} and \eqref{1st.order.vac.energy}, finding the Green's function expression for all regions $A1$, $A2$, $B1$ and $B2$ is required. Hence, the equation of motion was solved, and all the wave functions were computed. In the following, we apply the boundary condition to the wave functions. To apply the mixed boundary condition for region $A1$, the Dirichlet boundary condition is satisfied on the left line\,(\emph{e.g.}, the line placed on $x=-a/2$), and on the opposite side, the line placed on the right\,($x=+a/2$) satisfies the Neumann boundary condition. It is worth mentioning that the reverse order in applying the Dirichlet and Neumann boundary conditions on the lines will not change the vacuum energy expression. For the original region $A1$ displayed in fig.\,(\ref{BSS.Fig}), after performing the usual steps of the computation, the Green's function expression for the massive scalar field confined with mixed boundary condition between two lines located on $x = \pm a/2$ in two spatial dimensions after Wick rotation becomes\,\cite{peskin}:
\begin{eqnarray}\label{Green.Mixed.2D}
  G_{\hbox{\tiny{$\mathcal{M}$,$A1$}}}(r,r')=\frac{1}{a}\int\frac{d^2k}{(2\pi)^2}\sum_{n=0}^{\infty}\frac{e^{-\omega(t-t')}e^{ik_y(y-y')}[(-1)^n\cos(k_n x)+\sin(k_n x)][(-1)^n\cos(k_n x')+\sin(k_n x')]}{k^2+\omega_n^2},\nonumber\\
\end{eqnarray}
where $k=(\omega,k_y)$. Also, $k_n=\frac{(2n+1)\pi}{2a}$ and $\omega^2_n=k_n^2+m^2$ are the bounded wave-vector and wave-number, respectively. The subscript $\mathcal{M}$ denotes the type of the boundary condition imposed. Using eqs.\,(\ref{1st.order.vac.energy}) and (\ref{Green.Mixed.2D}) the total vacuum energy of region $A1$ per unit length is obtained as follows:
\begin{eqnarray}\label{vac.En.A1.1}
   E^{(1)}_{\hbox{\tiny{$\mathcal{M}$,$A1$}}}=\frac{-\lambda}{32\pi^2 a}\sum_{n=0}^{\infty}\sum_{n'=0}^{\infty}
 \bigg[\int_{0}^{\infty}\frac{KdK}{K^2+\omega^2_na^2}\bigg]\bigg[\int_{0}^{\infty}\frac{K'dK'}{K'^2+\omega^2_{n'}a^2}\bigg]\left(1+\frac{1}{2}\delta_{nn'}\right).
\end{eqnarray}
Both integrals over the nondimensionalized parameter $K=ka$ and $K'=k'a$ are logarithmic divergent due to their upper limits. To regularize them, we changed the upper limit of integrals by a cutoff, and expanded the results in the limit in which the cutoff goes to infinity. Therefore, we obtain:
\begin{eqnarray}\label{integral}
     \int_{0}^{\Lambda_{\mbox{\tiny$1(A1)$}}}\frac{KdK}{K^2+\omega_n^2a^2}=
     \frac{1}{2}\ln(1+\frac{\omega_n^2a^2}{\Lambda_{\mbox{\tiny$1(A1)$}}}^2)\buildrel{\Lambda_{\mbox{\tiny$1(A1)$}}\to\infty}
     \over\longrightarrow
     \ln\Lambda_{\mbox{\tiny$1(A1)$}}-\ln(\omega_na)-\mathcal{O}(\Lambda_{\mbox{\tiny$1(A1)$}}^{-2}).
\end{eqnarray}
This expansion for the integral result manifests the divergent parts. According to the definition of the Casimir energy via the BSS presented in eq.\,\eqref{BSS.Def.}, it is required to have the vacuum energy of each region shown in fig.\,(\ref{BSS.Fig}). Hence, given the Green's function expression for each region and using eq.\,\eqref{1st.order.vac.energy}, the vacuum energy of each region was obtained. In the following, using the BSS definition presented in eq.\,\eqref{BSS.Def.}, the total vacuum energies of two configurations $\mathcal{A}$ and $\mathcal{B}$ were subtracted from each other. Therefore, we have:
\begin{eqnarray}\label{vac.En.A1.2}
     \triangle E^{(1)}_{\hbox{\tiny{$\mathcal{M}$}}}&=&E^{(1)}_{\hbox{\tiny{$\mathcal{M}$,$\mathcal{A}$}}}-E^{(1)}_{\hbox{\tiny{$\mathcal{M}$,$\mathcal{B}$}}}\nonumber\\
     &=&\frac{-\lambda}{32\pi^2}\sum_{n,n'=0}^{\infty}
  \Bigg[\underbrace{\frac{1}{a}[\ln(\omega_na)-\ln\Lambda_{\mbox{\tiny$1(A1)$}}][\ln(\omega_{n'}a)-\ln\Lambda_{\mbox{\tiny$1(A1)$}}]}_{\mathcal{U}(a,\Lambda_{\mbox{\tiny$1(A1)$}})}
  +2\mathcal{U}\big(\frac{L-a}{2},\Lambda_{\mbox{\tiny$1(A2)$}}\big)-\{a\to b\}\Bigg]\left(1+\frac{1}{2}\delta_{nn'}\right),
\end{eqnarray}
A proper adjusting$^{1}$\footnotetext[1]{\footnotesize{\tiny} The relation for this adjustments can now be written as:
\begin{eqnarray}\label{adjust.cutoff.massless.0}
     \frac{\ln\Lambda_{\mbox{\tiny$1(A1)$}}}{\ln\Lambda_{\mbox{\tiny$1(B1)$}}}=\frac{\ln(\omega_nb)-\ln(\omega_{n'}b)-\ln\Lambda_{\mbox{\tiny$1(B1)$}}}{\ln(\omega_na)-\ln(\omega_{n'}a)-\ln\Lambda_{\mbox{\tiny$1(A1)$}}},
     \hspace{1.2cm}
     \frac{\ln\Lambda_{\mbox{\tiny$1(A2)$}}}{\ln\Lambda_{\mbox{\tiny$1(B2)$}}}=\frac{\ln(\omega_n(L-b))-\ln(\omega_{n'}(L-b))-\ln\Lambda_{\mbox{\tiny$1(B2)$}}}{\ln(\omega_n(L-a))
     -\ln(\omega_{n'}(L-a))-\ln\Lambda_{\mbox{\tiny$1(A2)$}}}.\nonumber
\end{eqnarray}}
for the cutoffs causes all infinite terms relevant to the cutoff to be eliminated, and the only remained terms from eq.\,\eqref{vac.En.A1.2} become:
\begin{eqnarray}\label{vac.En.A1.3}
     \triangle E^{(1)}_{\hbox{\tiny{$\mathcal{M}$}}}=\frac{-\lambda}{32\pi^2}\sum_{n=0}^{\infty}\sum_{n'=0}^{\infty}
     \Bigg[\underbrace{\frac{1}{a}\ln(\omega_na)\ln(\omega_{n'}a)}_{\mathcal{V}(a)}+2\mathcal{V}\big(\frac{L-a}{2}\big)-\{a\to b\}\Bigg]\left(1+\frac{1}{2}\delta_{nn'}\right),
\end{eqnarray}
Both summations over $n$ and $n'$ render eq.\,\eqref{vac.En.A1.3} to be divergent. To regularize them, we used the following form of Abel-Plana Summation Formula\,(APSF)\,\cite{Generalized.Abel.Plana.Saharian}:
\begin{eqnarray}\label{APSF}
   \sum_{n=1}^{\infty}f(n+\frac{1}{2})=\int_{0}^{\infty}f(x)dx-i\int_{0}^{\infty}\frac{f(it)-f(-it)}{e^{2\pi t}+1}dt.
\end{eqnarray}
This form of APSF is usually written for half-integer parameters and the first and second terms on the right hand side\,(rhs) of this summation formula are known as the \textit{integral term} and the \textit{branch-cut term}, respectively. The value of the integral term is usually obtained divergent, and on the contrary, the branch-cut term usually has a finite value. After applying the APSF to eq.\,\eqref{vac.En.A1.3}, we obtain:
\begin{eqnarray}\label{after.apsf.1}
    \triangle E^{(1)}_{\hbox{\tiny{$\mathcal{M}$}}}&=&\frac{-\lambda}{128\pi^2}\Bigg\{\frac{1}{a}\Big[\underbrace{\int_{0}^{\infty}\ln(x^2\pi^2+m^2a^2)dx}_{\mathcal{I}_1(a,\infty)}+\mathcal{B}_1(a)\Big]^2
     +\frac{1}{2a}\Big[\underbrace{\int_{0}^{\infty}\ln^2(x^2\pi^2+m^2a^2)dx}_{\mathcal{I}_2(a,\infty)}+\mathcal{B}_2(a)\Big]\nonumber\\
     &+&\frac{4}{L-a}\Big[\mathcal{I}_1\big(\frac{L-a}{2},\infty\big)+\mathcal{B}_1\big(\frac{L-a}{2}\big)\Big]^2
     +\frac{2}{L-a}\Big[\mathcal{I}_2\big(\frac{L-a}{2},\infty\big)+\mathcal{B}_2\big(\frac{L-a}{2}\big)\Big]\Bigg\}-\{a\to b\},
\end{eqnarray}
where $\mathcal{B}_1(\alpha)$ and $\mathcal{B}_2(\alpha)$ are the branch-cut terms of APSF. By expanding eq.\,\eqref{after.apsf.1}, we have:
\begin{eqnarray}\label{after.apsf.2}
     \triangle E^{(1)}_{\hbox{\tiny{$\mathcal{M}$}}}&=&\frac{-\lambda}{128\pi^2 }\Bigg\{\frac{1}{a}\Big[\mathcal{I}^2_1(a,\infty)+2\mathcal{I}_1(a,\infty)\mathcal{B}_1(a)+\mathcal{B}^2_1(a)
     +\frac{1}{2}\mathcal{I}_2(a,\infty)+\frac{1}{2}\mathcal{B}_2(a)\Big]+\frac{4}{L-a}\Big[\mathcal{I}^2_1\big(\frac{L-a}{2},\infty\big)\nonumber\\
     &+&2\mathcal{I}_1\big(\frac{L-a}{2},\infty\big)
     \mathcal{B}_1\big(\frac{L-a}{2}\big)
     +\mathcal{B}^2_1\big(\frac{L-a}{2}\big)+\frac{1}{2}\mathcal{I}_2\big(\frac{L-a}{2},\infty\big)
     +\frac{1}{2}\mathcal{B}_2\big(\frac{L-a}{2}\big)\Big]\Bigg\}-\{a\to b\},
\end{eqnarray}
where the integral terms $\mathcal{I}_1(\alpha,\infty)$ and $\mathcal{I}_2(\alpha,\infty)$ are divergent owing to their upper limits. To regularize them and remove their infinities, we used the cut-off regularization technique. The use of this technique supplemented by the BSS causes all infinities to be removed with less ambiguities from the integral terms $\mathcal{I}_1$ and $\mathcal{I}_2$. To present more details for the removal procedure, we start with the integral term $\mathcal{I}_2$ and, in the first step, we change the integration variable to $\xi=\frac{x\pi}{ma}$. Therefore, we obtain:
\begin{eqnarray}\label{I2}
     \mathcal{I}_2(a,\infty)&=&\int_{0}^{\infty}\Big(\ln(m^2a^2)+\ln\big(\frac{x^2\pi^2}{m^2a^2}+1\big)\Big)^2dx\nonumber\\
     &=&\frac{ma}{\pi}\Big[\int_{0}^{\infty}\ln^2(m^2a^2)d\xi+2\ln(m^2a^2)\int_{0}^{\infty}\ln(\xi^2+1)d\xi+\int_{0}^{\infty}\ln^2(\xi^2+1)d\xi\Big].
\end{eqnarray}
Since all integrals on the rhs of the above equation are divergent owing to their upper limits, we changed the upper limits of the first and second terms on the rhs of eq.\,\eqref{I2} to a cutoff value\,($\Lambda_{\mbox{\tiny$2(A1)$}}$). As we know, the term $\mathcal{I}_2(a,\infty)$ was produced from the vacuum energy of the region $A1$. A term similar to this term also appeared in the vacuum energy of the other regions in fig.\,(\ref{BSS.Fig}). Thus, we should choose a separate value for the cutoffs related to each region. As a result, the upper limits of the first and second integrals on the rhs of eq.\,\eqref{I2} for regions $A2$, $B1$, and $B2$ were replaced with $\Lambda_{\mbox{\tiny$2(A2)$}}$, $\Lambda_{\mbox{\tiny$2(B1)$}}$, and $\Lambda_{\mbox{\tiny$2(B2)$}}$, respectively. We maintain that if a proper value for the cutoffs is adjusted, supplemented by the subtraction procedure defined by BSS in eq.\,\eqref{after.apsf.2}, all infinities will be removed due to the first and second integrals on the rhs of eq.\,\eqref{I2}. It is worth mentioning that there is sufficient degree of freedom for these adjustments. Therefore, we adjust the cutoffs as follows:
\begin{eqnarray}\label{adjusting.cutoffs.}
     \frac{\Lambda_{\mbox{\tiny$2(A1)$}}}{\Lambda_{\mbox{\tiny$2(B1)$}}}&=&\frac{\ln(mb)}{\ln(ma)}\left[\frac{\ln(mb)+2\ln\Lambda_{\mbox{\tiny$2(B1)$}}-2+\frac{2\pi}{\Lambda_{\mbox{\tiny$2(B1)$}}}\ln(mb)}{\ln(ma)
     +2\ln\Lambda_{\mbox{\tiny$2(A1)$}}-2+\frac{2\pi}{\Lambda_{\mbox{\tiny$2(A1)$}}}\ln(ma)}\right],\nonumber\\
     \frac{\Lambda_{\mbox{\tiny$2(A2)$}}}{\Lambda_{\mbox{\tiny$2(B2)$}}}&=&\frac{\ln\big(\frac{m(L-b)}{2}\big)}{\ln\big(\frac{m(L-a)}{2}\big)}\left[\frac{\ln\big(\frac{m(L-b)}{2}\big)
     +2\ln\Lambda_{\mbox{\tiny$2(B2)$}}-2+\frac{2\pi}{\Lambda_{\mbox{\tiny$2(B2)$}}}\ln\big(\frac{m(L-b)}{2}\big)}
     {\ln\big(\frac{m(L-a)}{2}\big)+2\ln\Lambda_{\mbox{\tiny$2(A2)$}}-2+\frac{2\pi}{\Lambda_{\mbox{\tiny$2(A2)$}}}\ln\big(\frac{m(L-a)}{2}\big)}\right].
\end{eqnarray}
This adjustment for the cutoffs guarantees that all divergent contributions originated from the first and second terms of $\mathcal{I}_2$s are omitted from eq.\,\eqref{after.apsf.2}. For the third part on the rhs of eq.\,\eqref{I2}, after substituting $\mathcal{I}_2$s in eq.\,\eqref{after.apsf.2}, we obtain:
\begin{eqnarray}\label{I2-third}
     \frac{1}{a}\frac{ma}{2\pi}\int_{0}^{\infty}\ln^2(\xi^2+1)d\xi+\frac{4}{L-a}\frac{m(L-a)}{4\pi}\int_{0}^{\infty}\ln^2(\xi^2+1)d\xi-\{a\to b\}\nonumber\\
     =\Big[\frac{m}{2\pi}+\frac{m}{\pi}-\frac{m}{2\pi}-\frac{m}{\pi}\Big]\int_{0}^{\infty}\ln^2(\xi^2+1)d\xi=0.
\end{eqnarray}
In fact, when the subtraction process defined by BSS is applied, all parts of the third term on the rhs of eq.\,\eqref{I2} automatically cancel out each other. Hence, no contribution from the integral $\mathcal{I}_2$s remains in the final expression of the Casimir energy. Similar to what occurred in the removal procedure of the integral term $\mathcal{I}_2$, to regularize and remove infinities originated from the integral $\mathcal{I}_1$, is conducted again. Therefore, the BSS supplemented by the cutoff regularization technique is re-applied. For this purpose, the upper limit of the integral $\mathcal{I}_1$ was replaced with a cut-off value. For each distinct region in fig.\,(\ref{BSS.Fig}), the value of the cutoff was defined separately. Therefore, the values of $\Lambda_{\mbox{\tiny$3(A2)$}}$, $\Lambda_{\mbox{\tiny$3(B1)$}}$, and $\Lambda_{\mbox{\tiny$3(B2)$}}$ were replaced on the upper limits of the integral term $\mathcal{I}_1$ related to regions $A2$, $B1$, and $B2$, respectively. Then, to show the divergent part of integral $\mathcal{I}_1$, we calculated this integral up to the cutoff value and expanded the result at the infinite limit of cutoffs. This process led to finding an expansion as a function of the cutoffs. This expansion detaches the divergent contribution from the integral $\mathcal{I}_1$ clearly. This procedure converts the integral $\mathcal{I}_1(a,\Lambda_{\mbox{\tiny$3(A1)$}})$ to:
\begin{eqnarray}\label{I1(a)}
    &&\hspace{-2cm}\mathcal{I}_1(a,\Lambda_{\mbox{\tiny$3(A1)$}})=\frac{ma}{\pi}\ln(m^2a^2)\int_{0}^{\Lambda_{\mbox{\tiny$3(A1)$}}}d\xi+\frac{ma}{\pi}\int_{0}^{\Lambda_{\mbox{\tiny$3(A1)$}}}\ln(\xi^2+1)d\xi\nonumber\\
   && \buildrel {\Lambda_{\mbox{\tiny$3(A1)$}}\to\infty}\over \longrightarrow \frac{ma}{\pi}\Big[\ln(m^2a^2)\Lambda_{\mbox{\tiny$3(A1)$}}+2\Lambda_{\mbox{\tiny$3(A1)$}}(\ln\Lambda_{\mbox{\tiny$3(A1)$}}-1)+\pi-\Lambda_{\mbox{\tiny$3(A1)$}}^{-1}+\mathcal{O}(\Lambda_{\mbox{\tiny$3(A1)$}}^{-3})\Big],
\end{eqnarray}
where $\xi=\frac{x\pi}{ma}$. The term $\mathcal{I}_1(a,\Lambda_{\mbox{\tiny$3(A1)$}})$ was written for the region $A1$ of fig.\,(\ref{BSS.Fig}). 
Now, after substituting the integral result $\mathcal{I}_1$ for the second term in the bracket of eq.\,(\ref{after.apsf.2}), we obtain:
\begin{eqnarray}\label{2I1B1}
    &&\hspace{-2cm}\frac{2}{a}\mathcal{I}_1(a)\mathcal{B}_1(a)+\frac{4}{L-a}\mathcal{I}_1\big(\frac{L-a}{2}\big)\mathcal{B}_1\big(\frac{L-a}{2}\big)-\{a\to b\}\nonumber\\
    &&\hspace{0cm}=\frac{2m}{\pi}\mathcal{B}_1(a)\Big[\ln(m^2a^2)\Lambda_{\mbox{\tiny$3(A1)$}}+2\Lambda_{\mbox{\tiny$3(A1)$}}(\ln\Lambda_{\mbox{\tiny$3(A1)$}}-1)+\pi-\Lambda_{\mbox{\tiny$3(A1)$}}^{-1}+\mathcal{O}(\Lambda_{\mbox{\tiny$3(A1)$}}^{-3})\Big]\nonumber\\
    &&\hspace{0cm}+\frac{4m}{\pi}\mathcal{B}_1\big(\frac{L-a}{2}\big)\Big[\ln(\frac{m^2(L-a)^2}{4})\Lambda_{\mbox{\tiny$3(A2)$}}+2\Lambda_{\mbox{\tiny$3(A2)$}}(\ln\Lambda_{\mbox{\tiny$3(A2)$}}-1)+\pi-\Lambda_{\mbox{\tiny$3(A2)$}}^{-1}
    +\mathcal{O}(\Lambda_{\mbox{\tiny$3(A2)$}}^{-3})\Big]-\{a\to b\}\nonumber\\
    &&\buildrel {\mbox{ \tiny BSS}}\over\longrightarrow 2m\mathcal{B}_1(a)+4m\mathcal{B}_1\big(\frac{L-a}{2}\big)-\{a\to b\},
\end{eqnarray}
where in the last line in the above equation, the following adjustment for cutoffs was performed:
\begin{eqnarray}\label{cutoff.adjust.2I1B1}
      \frac{\Lambda_{\mbox{\tiny$3(A1)$}}}{\Lambda_{\mbox{\tiny$3(B1)$}}}=\frac{\mathcal{B}_1(b)}{\mathcal{B}_1(a)}\Bigg[\frac{\ln(mb)+\ln\Lambda_{\mbox{\tiny$3(B1)$}}-1}{\ln(ma)+\ln\Lambda_{\mbox{\tiny$3(A1)$}}-1}\Bigg],\hspace{1.5cm}
      \frac{\Lambda_{\mbox{\tiny$3(A2)$}}}{\Lambda_{\mbox{\tiny$3(B2)$}}}=\frac{\mathcal{B}_1\big(\frac{L-b}{2}\big)}{\mathcal{B}_1\big(\frac{L-a}{2}\big)}
      \Bigg[\frac{\ln\big(m\big(\frac{L-b}{2}\big)\big)+\ln\Lambda_{\mbox{\tiny$3(B2)$}}-1}{\ln\big(m\big(\frac{L-a}{2}\big)\big)+\ln\Lambda_{\mbox{\tiny$3(A2)$}}-1}\Bigg].
\end{eqnarray}
By conducting a similar scenario for the first term in the bracket of eq.\eqref{after.apsf.2}, it can be shown that all infinities would be removed due to this term, and there will not remain any contribution from this term in the final expression of the Casimir energy. Therefore, for eq.\,\eqref{after.apsf.2}, we have:
\begin{figure}[th] \hspace{0cm}\includegraphics[width=9.5cm]{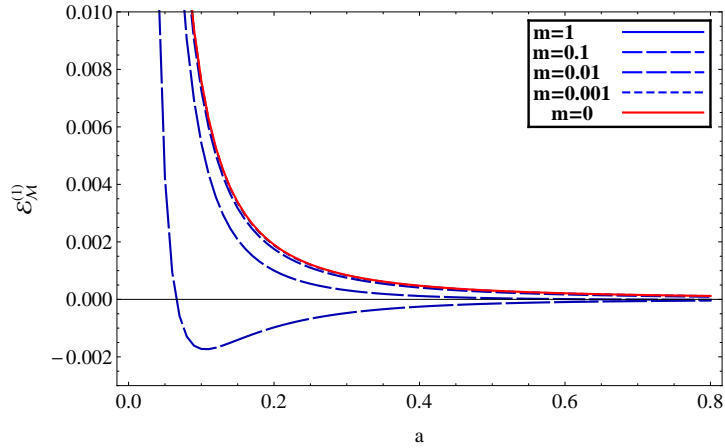}\hspace{0cm}
\caption{\label{Fig.2} \small
  The plot of the first-order radiative correction to the Casimir energy density of massive scalar fields between a pair of lines with distance $a$ for $\lambda=0.1$; the figure also presents the sequence of plots for the values of mass $m=\{1,0.1,0.01,0.001,0\}$. This sequence of plots shows that, when the mass parameter in the system is smaller, the Casimir energy value of the system for the massive case reaches the Casimir energy value for the massless case. This approaching trend is rapid and there is no remained significant difference for $m<0.001$. All units, including the distance of the plates $a$ and the Casimir energy, are considered in the natural unit\,($\hbar c=1$). }
\end{figure}
\begin{eqnarray}\label{after.apsf.3}
     \triangle E^{(1)}_{\hbox{\tiny{$\mathcal{M}$}}}&=&\frac{-\lambda}
     {128\pi^2 }\Bigg\{\frac{1}{a}\Big[2ma\mathcal{B}_1(a)+\mathcal{B}^2_1(a)+\frac{1}{2}\mathcal{B}_2(a)\Big]\nonumber\\
     &+&\frac{4}{L-a}\Big[m(L-a)\mathcal{B}_1\big(\frac{L-a}{2}\big)
     +\mathcal{B}^2_1\big(\frac{L-a}{2}\big)+\frac{1}{2}\mathcal{B}_2\big(\frac{L-a}{2}\big)\Big]\Bigg\}-\{a\to b\}.
\end{eqnarray}
In this equation, all the remaining terms are the branch-cut terms, and their values are finite. For the branch-cut term $\mathcal{B}_1(x)$, we have:
\begin{eqnarray}\label{B1}
     \mathcal{B}_1(x)=2mx\int_{1}^{\infty}\frac{d\eta}{e^{2mx\eta}+1}=\ln(1+e^{-2mx}),
\end{eqnarray}
and the value for the branch-cut term $\mathcal{B}_2(x)$ becomes:
\begin{eqnarray}\label{B2}
     \mathcal{B}_2(x)=4\pi\int_{\frac{mx}{\pi}}^{\infty}\frac{\ln(t^2\pi^2-m^2x^2)}{e^{2\pi t}+1}dt=4\ln(mx)\mathcal{B}_1(x)+4mx\int_{1}^{\infty}\frac{\ln(\eta^2-1)}{e^{2mx\eta}+1}d\eta,
\end{eqnarray}
where $\eta=\frac{\pi t}{mx}$. Finding a closed form for the second integral on the rhs of eq.\,\eqref{B2} is a difficult task. Hence, the denominator of the integrand was expanded and the result for the branch-cut term $\mathcal{B}_2(x)$ was obtained as:
\begin{eqnarray}\label{B2.final.result}
     \mathcal{B}_2(x)=4\ln(mx)\mathcal{B}_1(x)-2\sum_{j=1}^{\infty}\frac{e^{-2mxj}}{j}\Big[\gamma-e^{4mxj}\Gamma(0,4mxj)+\ln(mxj)\Big],
\end{eqnarray}
where the function $\Gamma(\alpha,y)$ is the incomplete Gamma function, which in our case, it is equal to:
\begin{eqnarray}\label{gamma.bessl.}
      \Gamma(0,y)=-e^{-y/2}\sqrt{\frac{y}{\pi}}\partial_\nu K_\nu(y/2)|_{\nu=-1/2}
\end{eqnarray}
In the final step, we computed all limits defined in eq.\,\eqref{BSS.Def.} for eq.\,\eqref{after.apsf.3}. Therefore, the radiative correction to the Casimir energy for massive scalar field confined with mixed boundary condition between two lines in $2+1$ dimensions per unit length is obtained as:
\begin{eqnarray}\label{Mixed.ECas}
    E^{(1)}_{\mathcal{M},\mbox{\tiny Cas.}}=\frac{-\lambda}{128\pi^2 a}\Big[2ma\mathcal{B}_1(a)+\mathcal{B}^2_1(a)+\frac{1}{2}\mathcal{B}_2(a)\Big].
\end{eqnarray}
This result is finite for any value of mass $m\neq0$ and $a\neq0$. An important extreme limit of the massive case of the Casimir energy is usually known as the massless limit. Since the branch-cut term $\mathcal{B}_2(a)$ in the limit $m\to0$ has a divergent value, the direct computation of the massless limit from eq.\,\eqref{Mixed.ECas} is not an easy task. Fortunately, the divergences appeared in the calculation of the massless limit are not essential and can be resolved. Thus, to resolve the divergence and find a physical and finite answer for the radiative correction to the Casimir energy for the massless case, we return to eq.\,\eqref{vac.En.A1.3} and set $m=0$. Therefore, we obtain:
\begin{eqnarray}\label{vac.En.massless}
     \triangle E^{(1)}_{\mbox{\tiny{$\mathcal{M}$}}}&=&E^{(1)}_{\hbox{\tiny{$\mathcal{M}$,$\mathcal{A}$}}}-E^{(1)}_{\hbox{\tiny{$\mathcal{M}$,$\mathcal{B}$}}}
     =\frac{-\lambda}{32\pi^2}\sum_{n,n'=0}^{\infty}
  \Bigg\{\frac{1}{a}\ln[(n+1/2)\pi]\ln[(n'+1/2)\pi]\nonumber\\
  &&\hspace{2.1cm}+\frac{4}{L-a}\ln[(n+1/2)\pi]\ln[(n'+1/2)\pi]-\{a\to b\}\Bigg\}\left(1+\frac{1}{2}\delta_{nn'}\right).
\end{eqnarray}
After applying the APSF given in eq.\,\eqref{APSF} to eq.\,\eqref{vac.En.massless}, we have:
\begin{eqnarray}\label{after.apsf.1.massless}
      \triangle E^{(1)}_{\hbox{\tiny{$\mathcal{M}$}}}
      &=&\frac{-\lambda}{32\pi^2 }\Bigg\{\frac{1}{a}\Big[\underbrace{\int_{0}^{\infty}\ln(x\pi)dx}_{\mathcal{J}_1(\infty)}+\mathcal{B}_1\Big]^2
     +\frac{1}{2a}\Big[\underbrace{\int_{0}^{\infty}\ln^2(x\pi)dx}_{\mathcal{J}_2(\infty)}+\mathcal{B}_2\Big]\nonumber\\
     &+&\frac{4}{L-a}\Big[\int_{0}^{\infty}\ln(x\pi)dx+\mathcal{B}_1\Big]^2
     +\frac{2}{L-a}\Big[\int_{0}^{\infty}\ln^2(x\pi)dx+\mathcal{B}_2\Big]\Bigg\}-\{a\to b\},
\end{eqnarray}
where $\mathcal{B}_1$ and $\mathcal{B}_2$ are the branch-cut terms of APSF, and their values are finite. Performing the calculation for them gives:
\begin{eqnarray}\label{B1&B2.massless}
     \mathcal{B}_1&=&-i\int_{0}^{\infty}\frac{\ln(it\pi)-\ln(-it\pi)}{e^{2\pi t}+1}dt=\pi\int_{0}^{\infty}\frac{dt}{e^{2\pi t}+1}=\frac{\ln2}{2},\nonumber\\
     \mathcal{B}_2&=&-i\int_{0}^{\infty}\frac{\ln^2(it\pi)-\ln^2(-it\pi)}{e^{2\pi t}+1}dt=2\pi\int_{0}^{\infty}\frac{\ln{(t\pi)}dt}{e^{2\pi t}+1}=\frac{-3}{2}\ln^22.
\end{eqnarray}
Two types of integral terms $\mathcal{J}_1(\infty)$ and $\mathcal{J}_2(\infty)$ shown in eq.\,\eqref{after.apsf.1.massless} are divergent. To remove their infinities by the subtraction process of the vacuum energy shown in eq.\,\eqref{after.apsf.1.massless}, we used the cutoff regularization technique again. Hence, we replaced the upper limits of integrals $\mathcal{J}_1$ and $\mathcal{J}_2$ with cutoffs $\Lambda_{A1}$, $\Lambda_{A2}$, $\Lambda_{B1}$, and $\Lambda_{B2}$ related to regions $A1$, $A2$, $B1$, and $B2$, respectively. Then, integrals $\mathcal{J}_1$ and $\mathcal{J}_2$ up to a finite value of the cutoff were computed, and the integral result was expanded in the infinite limit of the cutoffs:
\begin{eqnarray}\label{J1}
    \mathcal{J}_1(\Lambda)&=&\int_{0}^{\Lambda}\ln(x\pi)dx\buildrel {\Lambda\to\infty}\over\longrightarrow (\ln(\Lambda\pi)-1)\Lambda,\nonumber\\
    \mathcal{J}_2(\Lambda)&=&\int_{0}^{\Lambda}\ln^2(x\pi)dx\buildrel {\Lambda\to\infty}\over\longrightarrow \Big((\ln(\Lambda\pi)-1)^2+1\Big)\Lambda.
\end{eqnarray}
Afterward, the final expansions for each integral terms $\mathcal{J}_1$ and $\mathcal{J}_2$ given in eq.\,\eqref{J1} were substituted in eq.\,\eqref{after.apsf.1.massless}. We maintain that, the proper adjusting for the values of the cutoff supplemented by the subtraction procedure defined by the BSS will not leave any divergent contribution from integral terms $\mathcal{J}_1$ and $\mathcal{J}_2$.$^{1}$ As a result, in eq.\,\eqref{after.apsf.1.massless}, only the branch-cut terms remain, and finally by applying the limits $L/b\to\infty$ and $b/a\to\infty$, defined by eq.\,\eqref{BSS.Def.}, the final result for the radiative correction term to the Casimir energy per unit length for the massless scalar field confined with mixed boundary condition between two lines in two spatial dimensions becomes:
\footnotetext[1]{\footnotesize{\tiny} The relation for this adjustments can now be written as:
\begin{eqnarray}\label{adjust.cutoff.massless}
     \frac{\Lambda_{A1}}{\Lambda_{B1}}=\frac{a}{b}\bigg[\frac{3(\ln(\pi\Lambda_{B1})-1)^2+(\ln(\pi\Lambda_{B1})-1)\ln4+1}
     {3(\ln(\pi\Lambda_{A1})-1)^2+(\ln(\pi\Lambda_{A1})-1)\ln4+1}\bigg],\hspace{1.2cm}
     \frac{\Lambda_{A2}}{\Lambda_{B2}}=\frac{L-a}{L-b}\bigg[\frac{3(\ln(\pi\Lambda_{B2})-1)^2+(\ln(\pi\Lambda_{B2})-1)\ln4+1}
     {3(\ln(\pi\Lambda_{A2})-1)^2+(\ln(\pi\Lambda_{A2})-1)\ln4+1}\bigg].\nonumber
\end{eqnarray}}
\begin{eqnarray}\label{final.casimir.maassless}
     E^{(1)}_{\mathcal{M},\mbox{\tiny Cas.}}=\frac{-\lambda}{32\pi^2 a}\Big[\mathcal{B}^2_1(a)+\frac{1}{2}\mathcal{B}_2(a)\Big]=\frac{\lambda}{64\pi^2 a}\ln^22.
\end{eqnarray}
Fig.\,(\ref{Fig.2}) presents the consistency of results for massive and massless cases. This figure shows that the value of the first-order radiative correction to the Casimir energy reaches the values obtained from plot for the massless case, when the parameter $m$ goes to $0$. The obtained result for massive and massless cases are different from those reported in \cite{cavalcanti.1,cavalcanti.2}. It has to be noted that the main source of this difference is in the type of the counterterm employed in the renormalization program. In the previous work, to renormalize the bare parameter of the Lagrangian, free counterterms were used, while in this study, we used position-dependent counterterms. The position-dependent counterterms are consistent with dominant boundary conditions in the problem.
\par
\begin{figure}[th] \hspace{0cm}\includegraphics[width=9.5cm]{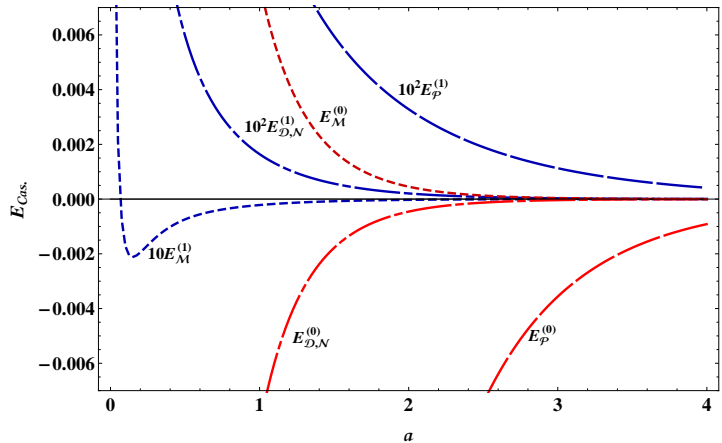}\hspace{0cm}
\caption{\label{Fig.3} \small
  The zero- and first-order radiative correction to the Casimir energy per unit length for the massive scalar field confined between two parallel lines by distance $a$ with four Dirichlet, Neumann, Periodic and mixed boundary conditions as a function of the distance of the lines is plotted. The superscript $(0)$ denotes the zero (or leading) order term of the Casimir energy and the superscript $(1)$ denotes the first-order term of the Casimir energy. The subscripts $\mathcal{D}$, $\mathcal{N}$, and $\mathcal{P}$ denote the type of the boundary condition, namely Dirichlet, Neumann, and Periodic boundary conditions. The values of the mass and coupling constant in all plots are considered $m=1$ and $\lambda=0.1$, respectively. All units including the mass of the field, the distance of the plates $a$, and the Casimir energy, are considered in the natural unit\,($\hbar c=1$).}
\end{figure}
\begin{figure}[th] \hspace{0cm}\includegraphics[width=9.5cm]{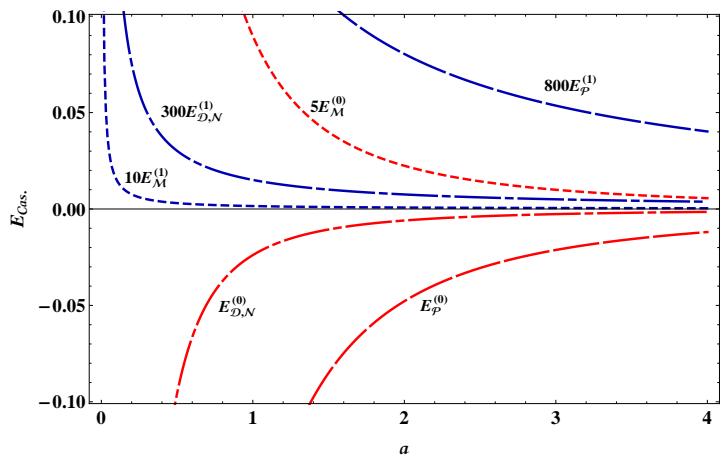}\hspace{0cm}
\caption{\label{Fig.4} \small
  The zero- and first-order radiative correction to the Casimir energy per unit length for the massless scalar field confined between two parallel lines by distance $a$ with four Dirichlet, Neumann, Periodic and mixed boundary conditions as a function of the distance of the lines is plotted. The superscripts $(0)$ denotes the zero (or leading) order term of the Casimir energy and the superscript $(1)$ denotes the first-order term of the Casimir energy. The subscript $\mathcal{D}$, $\mathcal{N}$, and $\mathcal{P}$ denote the type of the boundary condition, namely Dirichlet, Neumann, and Periodic boundary conditions. The value of the coupling constant in all plots is $\lambda=0.1$. All units including the mass of the field, the distance of the plates $a$, and the Casimir energy, are considered in the natural unit\,($\hbar c=1$).}
\end{figure}

The zero- and first-order corrections to the Casimir energy for massive and massless scalar fields with Dirichlet boundary condition confined between two lines in two spatial dimensions using the position-dependent counterterms were reported in\,\cite{2dim.valuyan}. By possessing the Dirichlet Casimir energy and using the following equation, the Casimir energy for the Neumann and periodic boundary conditions in any order of coupling constant $\lambda$ would be available,
\begin{eqnarray}\label{relation.periodic.Dirichlet.Neumann}
    E_{\mathcal{D}}(a)=E_{\mathcal{N}}(a)=\frac{1}{2}E_{\mathcal{P}}(2a),
\end{eqnarray}
where subscript $\mathcal{D}$, $\mathcal{N}$, and $\mathcal{P}$ denote the type of Dirichlet, Neumann, and Periodic boundary conditions, respectively. Therefore, based on the reported results for the Dirichlet Casimir energy and using eq.\,\eqref{relation.periodic.Dirichlet.Neumann}, the Casimir energy for Neumann and Periodic boundary conditions in both orders of corrections was extracted. Figs.\,(\ref{Fig.3}) and (\ref{Fig.4}) present the plot of the new results of this paper under mixed boundary conditions along with the obtained results for the zero- and first-orders of correction to the Casimir energy under Dirichlet, Neumann, Periodic boundary conditions extracted from eq.\,\eqref{relation.periodic.Dirichlet.Neumann}. Fig.\,(\ref{Fig.3}) displays the Casimir energy for both orders of correction per unit length as a function of the distance of the lines ($a$) for the massive scalar field. In Fig.\,(\ref{Fig.4}) these quantities for the massless case were plotted. These figures provide us with an opportunity to compare the sign and magnitude of the Casimir energy of the scalar field to four types of boundary conditions with each other. In fig.\,(\ref{Fig.3}), all plots except for the Casimir energy with mixed boundary condition\,($E^{(1)}_{\mathcal{M}}$), by increasing the distance between the lines, the value of the Casimir energy decreases. Whereas, the first-order Casimir energy of mixed boundary condition has a minimum in its graph. This minimum shows that the Casimir force in the first-order correction(which is defined as the derivative of the energy with respect to the distance of lines) changes the sign at a specific distance of lines.
\section*{3. Conclusion}\label{sec:conclusion}
In the present work, the zero- and first-order radiative correction to the Casimir energy was computed for massive and massless scalar field confined with two lines with mixed\,(Dirichlet-Neumann) boundary conditions. The main difference between our work and those reported previously\,\cite{cavalcanti.1,cavalcanti.2} is in the details of the renormalization program used. In the literature, to renormalize the bare parameters of the Lagrangian for any problem designed in the Casimir energy subject, the free counterterms are usually used. While, in this study, the position-dependent counterterms were employed. The position-dependent counterterms allow all influences originated from the boundary condition to be imported in the renormalization procedure. It makes the renormalization program to be in a self-consistent manner. Owing to this difference in the renormalization procedure, our results for the radiative correction to the Casimir energy differ from the ones reported previously. Our results also indicate that the sign of the first-order correction of the Casimir force for the massive scalar field with mixed boundary condition is changed at a specific distance of the lines. However, our results are in agreement with the expected physical basis and limits. Another noteworthy point in this study is to apply the Box Subtraction Scheme\,(BSS) as a regularization technique. The appearance of the logarithmic divergent expression is common for the vacuum energy in the problem designed in two spatial dimensions. However, the BSS supplemented by the cutoff regularization technique successfully regularized and removed these sorts of infinity without using any analytic continuation technique.
\acknowledgments
The Author would like to thank the research office of Semnan Branch, Islamic Azad University for financial support.

\end{document}